\documentclass[prl,twocolumn,showpacs,preprintnumbers,amsmath,amssymb]{revtex4}
\usepackage{graphicx}% Include figure files
\usepackage{dcolumn}% Align table columns on decimal point
\usepackage{bm}% bold math
\begin{document}

\preprint{APS/123-QED}

\title{Nernst effect in the phase-fluctuating superconductor InO$_x$}

\author{P. Spathis}
\email{Panayotis.Spathis@espci.Fr}
\author{H. Aubin}
\email{Herve.Aubin@espci.Fr}
\author{A. Pourret}
\author{K. Behnia}
\affiliation{Laboratoire Photons et Mati\`{e}re (CNRS), ESPCI, 10
rue Vauquelin, 75231 Paris, France}

\date{\today}% It is always \today, today,

\begin{abstract}

We present a study of the Nernst effect in amorphous 2D
superconductor InO$_x$, whose low carrier density implies low phase
rigidity and strong superconducting phase fluctuations. Instead of
presenting the abrupt jump expected at a BCS transition, the Nernst
signal evolves continuously through the superconducting transition
as previously observed in underdoped cuprates. This contrasts with
the case of Nb$_{0.15}$Si$_{0.85}$, where the Nernst signal due to
vortices below T$_{c}$ and by Gaussian fluctuations above are
clearly distinct. The behavior of the ghost critical field in
InO$_x$ points to a  correlation length which does not diverge at
$T_c$, a temperature below which the amplitude fluctuations freeze,
but phase fluctuations survive.

\end{abstract}

\maketitle

Those past years have witnessed the emergence of the Nernst effect
as an important probe of Superconducting Fluctuations (SF),
following the observation of an anomalous Nernst signal above T$_c$
in cuprates~\cite{NernstHTc}. In amorphous superconducting thin
films of Nb$_{0.15}$Si$_{0.85}$, a Nernst signal produced by
Cooper-pair fluctuations could be detected in a wide temperature and
field range ~\cite{pourret2006,pourret2007}. Close to $T_{c}$, the
magnitude of the Nernst coefficient found in this experiment was in
very good agreement with the predictions of a theory by Ussishkin,
Sondhi and Huse (USH) for the transverse thermoelectric response of
the Gaussian fluctuations of the Superconducting Order
Parameter(SOP)~\cite{huse2002}. This is not the case of underdoped
cuprates, where the Nernst signal does not follow the predictions of
the USH theory\cite{huse2002} and phase fluctuations of the SOP are
believed to play a major role.

To address this issue, new theories have been proposed addressing
cases where the Nernst signal is only generated by phase
fluctuations of the SOP~\cite{podolsky} or by quantum fluctuations
near a Superconductor-Insulator Transition
(SIT)~\cite{hartnoll-2007}. On the experimental side, recent
measurements on organic quasi-2D superconductors ~\cite{ardavan}
detected a finite Nernst signal above T$_{c}$ in a temperature range
widening with the approach of the Mott insulator as in the case of
cuprates~\cite{NernstHTc}. However, since the Nernst response of
normal electrons scales with their mobility ~\cite{behnia}, the
normal-state Nernst response is not negligible in either cuprate or
organic superconductors. This complicates any quantitative
comparison of the measured Nernst signal with theoretical
predictions.

\begin{figure}[h!]
\begin{center}
\includegraphics[width=9cm,keepaspectratio]{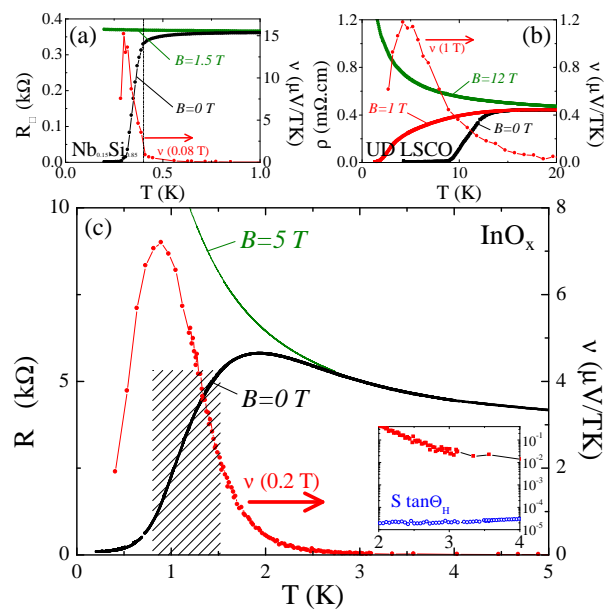}
\caption{\label{fig1} The effect of superconducting transition on
resistance and Nernst signal in Nb$_{0.15}$Si$_{0.85}$
~\cite{pourret2006}(panel $a$), La$_{1.94}$Sr$_{0.06}$CuO$_4$ single
crystal ~\cite{capan}(panel $b$) and InO$_x$ (panel $c$). The shaded
region represents the temperature range corresponding to 0.1-0.9 of
$R_\square^{max}$. The inset of panel $c$ compares the Nernst signal
with $S \tan \theta_{H}$. Note the sharp increase of the Nernst
signal at $T_c$ for Nb$_{0.15}$Si$_{0.85}$, in contrast to
continuous change observed for LSCO and InO$_{x}$.}
\end{center}
\end{figure}

In this Letter, we report on the case of InO$_x$. Several factors
make thin films of this system an appealing candidate for the study
of the Nernst signal generated by superconducting phase
fluctuations. First of all, due to its low carrier density, a poor
superfluid stiffness and, consequently, strong phase fluctuations
are expected~\cite{EMERY1995}. Moreover, the normal state is a
simple dirty metal, with a negligible Nernst response. This system
is also believed to host a Kosterlitz-Thouless-Berezinskii (KTB)
transition~\cite{crane}. Finally, due to its large sheet resistance,
quantum fluctuations of the phase of the SOP are expected to give
rise to a SIT at zero-temperature.

According to our findings, the Nernst effect in this system shares
common features with cuprates. In contrast with
Nb$_{0.15}$Si$_{0.85}$, its temperature dependence does not follow
the predictions of USH theory. Moreover, both the field and
temperature dependence of the Nernst signal in InO$_x$ indicate that
the blurred transition reflects a regime of superconducting
fluctuations whose Correlation Length(CL) does not diverge. Our
analysis is based on the previous study of the Nernst data in
Nb$_{0.15}$Si$_{0.85}$~\cite{pourret2006,pourret2007}, which
established the link between the  Nernst signal and the
CL~\cite{pourret2007}.

The 300 \AA{}-thick amorphous InO$_x$ film used in this study is
deposited on a glass substrate by $e$-gun evaporation of In$_2$O$_3$
in oxygen atmosphere~\cite{ovadyahu93}. Using a
one-heater-two-thermometers setup, four point resistance, Hall
effect and thermoelectric measurements are measured in a single
cool-down. The as-prepared film is insulating down to the lowest
measured temperature of 60~mK. After thermal annealing at
$50^\circ$C under vacuum as described elsewhere~\cite{Ovad}, the
sheet resistance decreases by about 30 \% and a superconducting
state appears. According to optical absorption experiments, this
drop of resistivity is the consequence of the volume shrinkage of
the sample during annealing~\cite{ovadyahu93}. During all
measurements, the film has been kept below liquid nitrogen
temperature to avoid aging effects.

Fig.~\ref{fig1} compares the behavior of the Nernst signal
$N=E_y/(-\nabla_x T)$, measured in the low field limit, in  the
vicinity of the superconducting transition in three different
systems. In the case of Nb$_{0.15}$Si$_{0.85}$~\cite{pourret2006},
$N$ increases abruptly at the BCS superconducting transition. It was
shown~\cite{pourret2006,pourret2007} that above $T_{c}$, the Nernst
signal is generated by Cooper pairs fluctuations, and below $T_{c}$,
by well defined vortices. In contrast, in
La$_{1.94}$Sr$_{0.06}$CuO$_4$ ~\cite{capan}, as seen in
Fig.~\ref{fig1}b, no distinct anomaly in $N(T)$ is visible at any
temperature separating these two regimes. As seen in
Fig.~\ref{fig1}c, the same is true for InO$_x$: The Nernst signal
evolves continuously across the superconducting transition. As seen
in the inset of the same figure, the signal is at least $100$ times
larger than the product of the Seebeck coefficient and the Hall
angle. Since the latter ($S\tan\theta$) sets the order of magnitude
of the normal-state response, the observed Nernst signal is almost
entirely due to SF.

A low carrier density is one fundamental feature shared by InO$_x$
and La$_{1.94}$Sr$_{0.06}$CuO$_4$. The Hall coefficient measured in
our film ($R_H=6.10^{-9}$ m$^3$.C$^{-1}$) is close to the one found
in La$_{1-x}$Sr$_{x}$CuO$_4$ (LSCO) at x=0.05~\cite{LSCO} and yields
a carrier density of $n=10^{21}$~cm$^{-3}$. On the other hand, the
Hall coefficient in Nb$_{0.15}$Si$_{0.85}$ is 80 times
lower~\cite{pourret2006}, implying a much higher carrier density.
Since the superfluid stiffness is proportional to the superfluid
density, superconductors such as InO$_x$ and
La$_{1.94}$Sr$_{0.06}$CuO$_4$ are expected to display strong
phase-fluctuations~\cite{EMERY1995}. This is the most plausible
source of this peculiar Nernst response in the vicinity of the
superconducting transition.

\begin{figure}[h!]
\begin{center}
\includegraphics[width=9cm,keepaspectratio]{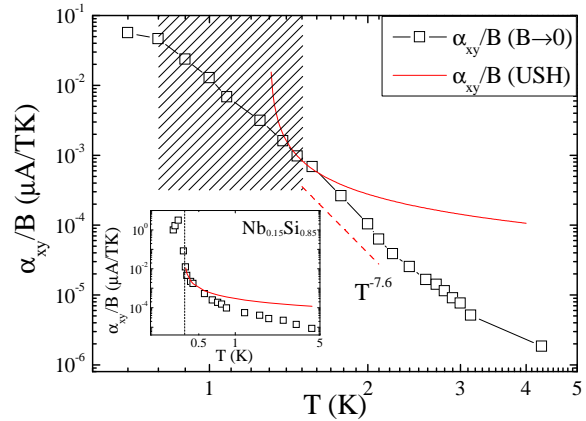}
\caption{\label{fig2} Transverse Peltier coefficient $\alpha_{xy}/B$
versus temperature for $B\rightarrow0$. The shaded region represents
the temperature range where $T_c$ is expected. The theoretical
prediction of USH theory is represented by the red line. The inset
shows data for Nb$_{0.15}$Si$_{0.85}$ along with the USH prediction
(from~\cite{pourret2006}). In this compound, $T_c=0.38$~K, is
represented by the vertical line.}
\end{center}
\end{figure}

Since $ \tan \theta_{H} \ll 1$ ,the Nernst and resistivity data
suffice to determine the transverse Peltier response
$\alpha_{xy}=N/R_\square$ as presented in Figure~\ref{fig2}. Above
$T_c$, for short-lived Cooper pairs described as Gaussian
fluctuations of the SOP, this coefficient is simply related to the
superconducting CL, $\alpha_{xy}/B\propto \xi^2$ at
2D~\cite{huse2002}. The inset of Figure~\ref{fig2} shows that in
Nb$_{0.15}$Si$_{0.85}$ cooling leads to a steep increase in
$(\frac{\alpha_{xy}}{B})_{B\rightarrow 0}$  at $T_c$, indicating the
divergence of the CL. In contrast, for InO$_x$, $\alpha_{xy}/B$
evolves continuously and no abrupt change is observed on the whole
temperature range of measurements; i.e; 0.6~K to 4.5~K. This
suggests that there is no diverging CL and therefore, no true phase
transition at $T_c$, the temperature corresponding to the formation
of Cooper pairs, expected to be located in the $0.8-1.5$~K range and
represented by shaded regions in figures~\ref{fig1} and ~\ref{fig2}.

\begin{figure}[h!]
\begin{center}
\includegraphics[width=9cm,keepaspectratio]{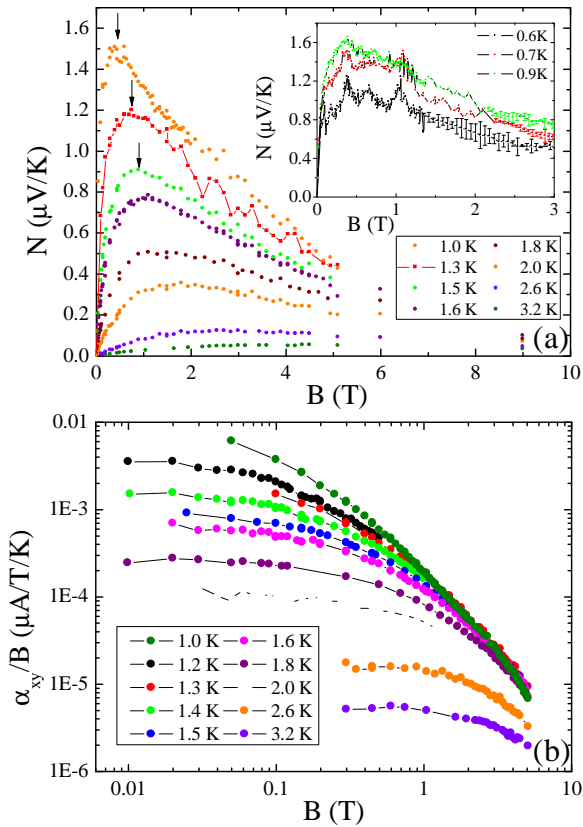}
\caption{\label{fig3}a) Nernst signal as a function of magnetic
field for different temperatures $T\geq 1~K$. The inset shows the
low temperature data, $T< 1~K$. Arrows indicate the ghost critical
field $B^\ast$. b) $\alpha_{xy}/B$ versus magnetic field in the
regime of superconducting fluctuations, $T>1~K$}
\end{center}
\end{figure}

We now proceed to an analysis of the field dependence of the Nernst
data, which leads to the same conclusion. Fig.~\ref{fig3}a shows
that, for each temperature, the Nernst signal $N(B)$ peaks with a
maximum at a temperature-dependent magnetic field scale $B^\ast(T)$.
This peak can be clearly observed in $N(B)$ down to a temperature of
0.9~K. As discussed in previous studies on
Nb$_{0.15}$Si$_{0.85}$~\cite{pourret2006,pourret2007}, at any
temperature and magnetic field, $\alpha_{xy}/B$ depends only on the
size of SF. At zero magnetic field, this size is set by the CL. At
high field, this size is set by the magnetic length
$l_B=(\hbar/2eB)^{1/2}$ when it becomes shorter than the zero-field
CL. Thus, this coefficient acquires a characteristic
field-temperature dependence that is observed in
Nb$_{0.15}$Si$_{0.85}$~\cite{pourret2007} and in InO$_{x}$ as shown
Fig.~\ref{fig3}b. $\alpha^{sc}_{xy}/B$ is field-independent at low
magnetic field, however, at high magnetic field, all the data evolve
toward a single curve weakly dependent on temperature. This
crossover is responsible for the peak observed at $B^\ast(T)$ in the
field dependence of the Nernst signal $N(B)$ (see arrows in
Fig.~\ref{fig3}a).

\begin{figure}[h!]
\begin{center}
\includegraphics[width=9cm,keepaspectratio]{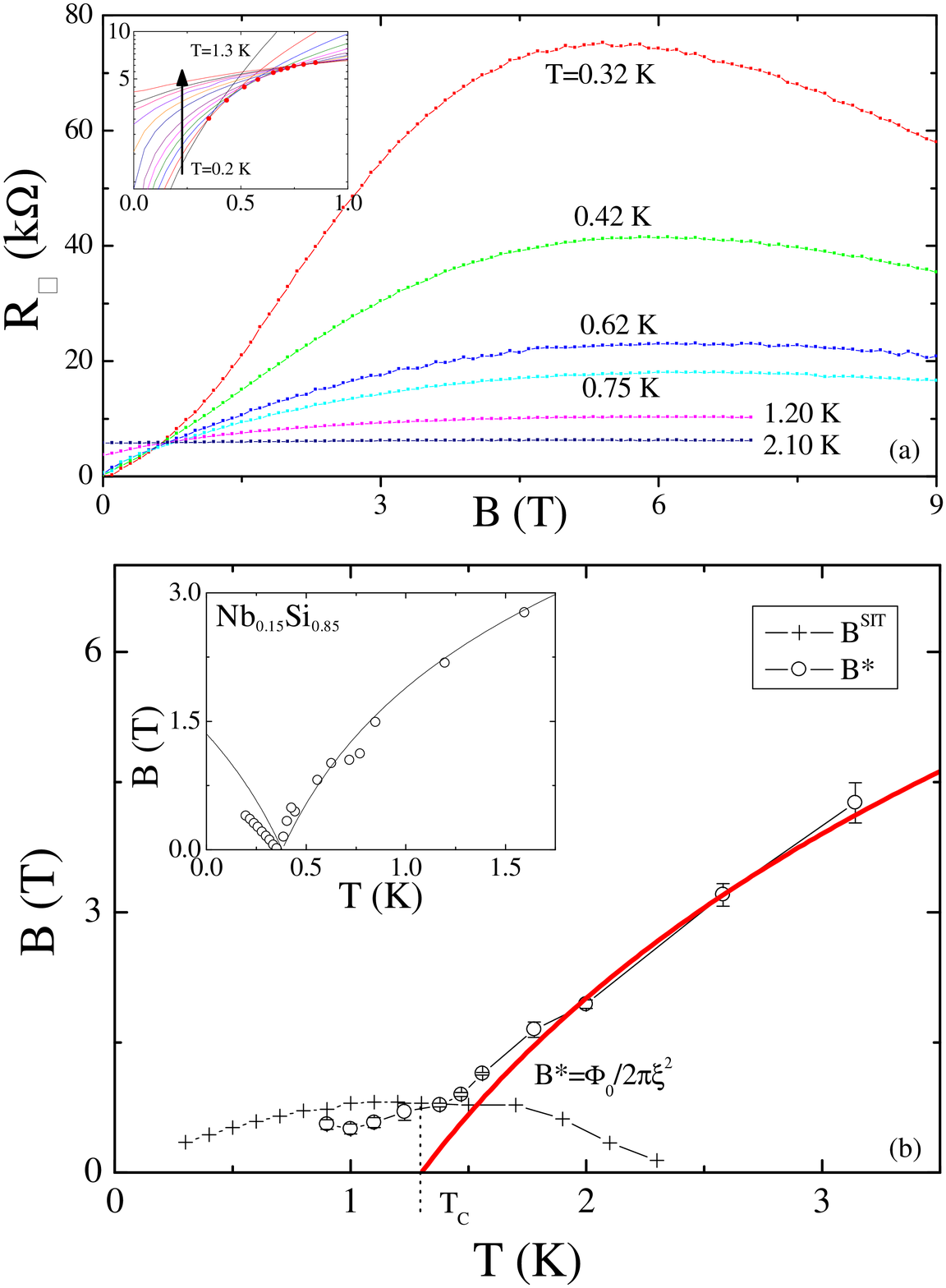}
\caption{ \label{fig4} a) Magnetoresistance measured for several
temperatures. The inset focus on the low field part showing the
crossing between adjacent isotherms. b) Phase diagram representing
$B_{SIT}$ and the position of the Nernst maximum $B^\ast$. The thick
line is an adjustment of the ghost critical field
$\Phi_0/2\pi\xi^2$. The inset shows $B^\ast$ measured in
Nb$_{0.15}$Si$_{0.85}$~\cite{pourret2007}. The line below $T_c$
represents $H_{c2}(T)$.}
\end{center}
\end{figure}

The temperature dependence of $B^\ast(T)$ is presented in fig.
\ref{fig4}b for InO$_{x}$ (main panel) and for
Nb$_{0.15}$Si$_{0.85}$~\cite{pourret2007} (inset). In both systems,
above $T_c$, $B^\ast\propto \ln T$ as expected for the Ghost
Critical Field (GCF), $\Phi_0/2\pi\xi^2$, set by the BCS CL,
$\xi=\xi_0\varepsilon^{-1/2}$ where $\varepsilon=\ln T/T_c$. In the
case of Nb$_{0.15}$Si$_{0.85}$ , $\xi_0$ and $T_c$ could be
independently determined and compared with the GCF determined from
the Nernst data. For InO$_{x}$,
 on the other hand, we set $T_c$ and $\xi_0$ such that the GCF
line fits $B^\ast(T)$. The thick line Fig.~\ref{fig4} is a fit using
$\xi_0=8.4\pm0.2$~nm and $T_c=1.2$~K. This value of $T_c$
corresponds to the mid-point of the resistive transition, as seen
Figure~\ref{fig1}. A similar conclusion on the position of $T_c$~
was drawn by a recent study on InO$_{x}$\cite{crane}.

With the temperature dependence of the CL just determined, we find
that USH formula\cite{huse2002}, when $B\rightarrow0$:

\begin{equation}
\alpha^{sc}_{xy}=\frac{1}{12\pi}\frac{k_B e}{\hbar}
\frac{\xi^2}{l_B^2}\label{eq:USH}
\end{equation}
is close to the measured $\alpha_{xy}$, as seen Fig.~\ref{fig2}.
However, $\alpha_{xy}$ decreases with temperature as fast as
$T^{-7.6}$, much faster than predicted by the USH theory. One
possibility is that the CL is too short for the applicability of the
USH theory on a large temperature range. Indeed, in
Nb$_{0.15}$Si$_{0.85}$, where $\xi_0$ is larger, $\alpha_{xy}/B$ was
found to deviate from USH theory for $T>1.3*T_c$, see inset of
Fig.\ref{fig2}. Since $\xi_0$ is shorter in InO$_x$, the deviation
from theory is expected to occur closer to T$_{c}$. Another
possibility is a deep difference in the nature of fluctuations in
the two systems. A recent model of phase-only
fluctuations~\cite{podolsky} predicts a faster decrease of the
Nernst signal above $T_{KTB}$ compared to what is expected in the
Gaussian picture in the temperature range above $T_{BCS}$, in
qualitative agreement with what is seen here. However, if the fast
decrease of the Nernst signal observed up to 4~K is due to
fluctuations with frozen amplitude, it would imply $T_{BCS}$ to be
above 4~K, which is unlikely.

While we find difficult to draw conclusions from temperature
dependence of the Nernst data, the interpretation of the field
position of the Nernst peak as the GCF appears straightforward.
According to our analysis, this field scale reflects the CL, no
matter the precise nature of SF, Gaussian or phase-only. This
recently received some theoretical support. Functional forms for the
field dependence compatible with a maximum at the GCF have been
predicted by a theory expanding the USH theory to finite
field~\cite{varlamov} and by a recent theory of the Nernst effect in
the vicinity of a SIT~\cite{hartnoll-2007}.

In Nb$_{0.15}$Si$_{0.85}$ (see inset of Fig.~\ref{fig4}b) the GCF
vanishes and at T$_{c}$ (reflecting the divergence of the CL)
mirrors the behavior of the H$_{c2}$(T) below T$_{c}$. One striking
observation of this work is the breakdown of this picture in InO$_x$
. As seen in Fig.~\ref{fig3}a, $B^\ast(T)$ keeps decreasing down to
$0.9$~K, well below our estimate of $T_c=1.2$~K. This indicates that
the CL does not diverge and that no true phase transition occurs at
$T_c$, the temperature associated with the formation of Cooper
pairs. An identical conclusion was drawn from the temperature
dependence of $\alpha_{xy}/B$ where no abrupt change is observed
upon crossing $T_c$. This leads us to conclude that the wide
superconducting transition  is not simply the consequence of a large
critical region, or sample inhomogeneity, but reflects the presence
of an intermediate fluctuation region between $T_{c}$, where
amplitude fluctuations freeze, and a lower temperature, $T_{KTB}$,
where phase coherence should be established. Such a region of
phase-only fluctuations in InO$_x$ was recently inferred from
high-frequency conductivity measurements~\cite{crane}.

Upon cooling,  phase fluctuations are expected to disappear at KTB
transition where the vortex and anti-vortex bind together. Since the
vortices are a major source of the Nernst signal, the latter should
be strongly affected by KTB transition. Below 0.9 K, the overall
magnitude of the Nernst signal decreases and the field dependence
$N(B)$ displays a broad maximum with complicated but reproducible
substructures. The reduced amplitude points to a reduced vortex
mobility below 0.9 K, a possible signature of the KTB transition.
The multiple peaks observed in $N(B)$ are reminiscent of what was
also observed in the field dependence of the Nernst signal in
hole-doped cuprates at low temperatures and tentatively attributed
to a plastic flow of vortices~\cite{NernstHTc}.

This behavior may also be related to the SIT and Bose insulating
properties of this system. As previously observed in
InO$_x$~\cite{Gantmakher2000,shahar}, at low temperature, the
magnetoresistance increases steeply following the magnetic-field
induced SIT, (see Fig.~\ref{fig4}a). According to the dirty-boson
model~\cite{Fisher90-prl65}, the insulating side is formed of Cooper
pairs localized by the quantum melting of the vortex system. Within
this framework, the negative magnetoresistance observed at high
field is due to the pair-breaking effect of the magnetic field when
$H>H_{c2}$. Using our previous estimation of the CL, we find
$H_{c2}=4.7\pm0.2 T$, which is about the position of the maximum of
the magnetoresistance curves, thus providing support to this
interpretation of the negative magnetoresistance.

In contrast to the SIT observed in
Nb$_{0.15}$Si$_{0.85}$~\cite{Aubin2006}, MoGe~\cite{moge}, or
Bi~\cite{goldman} thin films, the crossing-point $B_{SIT}(T)$,
reported Fig.~\ref{fig4}, is temperature dependent. This behavior
has been discussed previously for this
compound~\cite{Gantmakher2001} and remains yet to be understood.

To summarize, measuring Nernst signal and resistivity in InO$_x$, we
found that the transverse Peltier coefficient evolves continuously
across the superconducting transition. Furthermore, we find that the
GCF keeps decreasing on the temperature range where the Cooper pair
formation is expected to occur. This indicates that no true phase
transition occurs at $T_c$ and implies the existence of a regime of
phase-only fluctuations of the SOP. The similarity between the
temperature dependence of the Nernst signal measured in $InO_x$ and
the underdoped cuprates is additional support for the existence of a
regime of phase-only fluctuations in the latter system.

We are grateful to Z. Ovadyahu for the realization of InO$_x$
samples and C. Capan for the Nernst data on LSCO. The authors are
also grateful to M. Feigelman, M. Mueller, A. Vishwanath, S.
Sachdev, M. Skvortsov and A. Varlamov for discussions.

\bibliography{Inox}

\bibliographystyle{apsrev}

\end{document}